\documentclass[12pt]{article}
\usepackage{graphicx}

\def\pbnr{}
\def\speaker{Linlin Zhang}
\def\onbehalfof{the CMS collaboration}
\def\title{Measurements of Quarkonium Production and Polarization at CMS}
\def\affiliation{School of Physics \\
Peking University, Beijing, China}

\newcommand\pubnumber{\pbnr}
\newcommand\pubdate{\today}
\def\Title#1{\begin{center} {\Large #1 } \end{center}}
\def\Author#1{\begin{center}{ \sc #1} \end{center}}

\newcommand{\OnBehalf}[1]{\sbox0{#1}\ifdim\wd0=0pt
        {}
	\else
	{\\on behalf of #1}
	\fi}
\newcommand{\SupportedBy}[1]{\sbox0{#1}\ifdim\wd0=0pt
        {}
	\else
	{\footnote{#1}}
	\fi}
\def\Address#1{\begin{center}{ \it #1} \end{center}}

\newcommand\pubblock{\includegraphics[width=5cm]{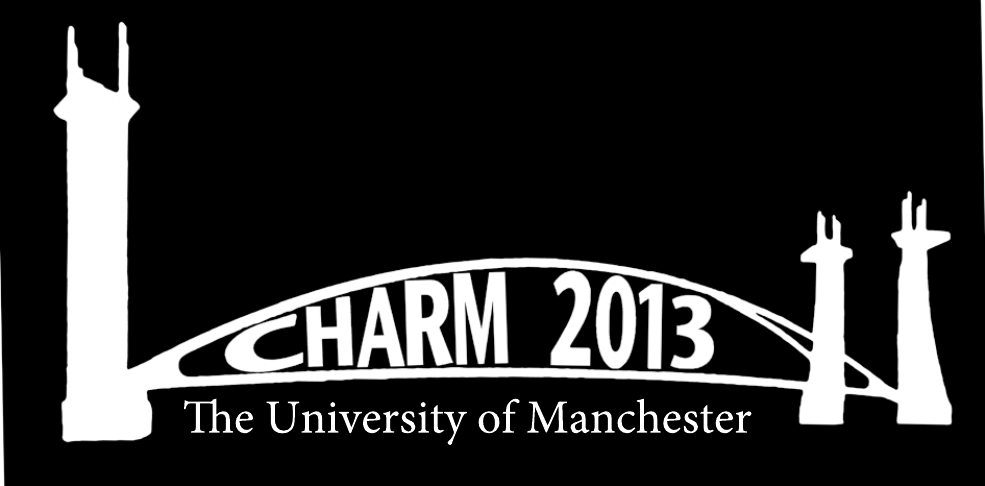}\hfill{\begin{tabular}{l} \pubnumber\\
         \pubdate  \end{tabular}}}
\newenvironment{Abstract}{\begin{quotation}  }{\end{quotation}}
\newenvironment{Presented}{\begin{quotation} \begin{center} 
             PRESENTED AT\end{center}\bigskip 
      \begin{center}\begin{large}}{\end{large}\end{center} \end{quotation}}

\def\venue{The 6$^{th}$ International Workshop on Charm Physics\\
(CHARM 2013)\\
Manchester, UK,  31 August -- 4 September, 2013}

\def\beq{\begin{equation}}
\def\eeq#1{\label{#1}\end{equation}}
\def\eeqn{\end{equation}}

\def\beqa{\begin{eqnarray}}
\def\eeqa#1{\label{#1}\end{eqnarray}}
\def\eeqan{\end{eqnarray}}

\let\bar=\overbar

\def\Dslash{\not{\hbox{\kern-4pt $D$}}}
\def\dslash{\not{\hbox{\kern-2pt $\del$}}}

\def\msb{{\bar{\ssstyle M \kern -1pt S}}}

\newcommand{\pT}{p_\mathrm{T}}
\newcommand{\UpsOne}{\Upsilon(1S)}
\newcommand{\UpsTwo}{\Upsilon(2S)}
\newcommand{\UpsThree}{\Upsilon(3S)}
\newcommand{\UpsNS}{\Upsilon(nS)}
\newcommand{\Jpsi}{\mathrm{J}/\psi}

\newcommand{\PsiTwo}{\psi(2S)}
\newcommand{\PsiNS}{\psi(nS)}

\newcommand{\costh}{\cos\vartheta}

\newcommand{\lamth}{\lambda_\vartheta}
\newcommand{\lamph}{\lambda_\varphi}
\newcommand{\lamthph}{\lambda_{\vartheta\varphi}}
\newcommand{\lamtilde}{\tilde{\lambda}}

\begin{document}
\begin{titlepage}
\pubblock

\vfill
\Title{\title}
\vfill
\Author{\speaker
\OnBehalf{\onbehalfof}}
\Address{\affiliation}
\vfill
\begin{Abstract}

The polarizations of $\UpsNS$ (n=1,2,3) and prompt $\Jpsi$ and $\PsiTwo$, as well as the differential cross section of the $\UpsNS$, are measured in proton-proton collisions at sqrt{s} = 7 TeV, using a dimuon data sample collected by the CMS experiment at the LHC, corresponding to an integrated luminosity of 4.9 $fb^{-1}$. The differential cross section is measured as a function of transverse momentum of $\UpsNS$. The data show a transition from exponential to power-law behavior in the neighborhood of 20 GeV, and the power-law exponents for all three states are consistent. The polarization parameters $\lamth$, $\lamph$, and $\lamthph$, as well as the frame-invariant quantity $\lamtilde$, are measured from the dimuon decay angular distributions in three different polarization frames. No evidence of large polarizations is seen in these kinematic regions, which extend much beyond those previously explored.

\end{Abstract}
\vfill
\begin{Presented}
\venue
\end{Presented}
\vfill
\end{titlepage}
\def\thefootnote{\fnsymbol{footnote}}
\setcounter{footnote}{0}


\section{Introduction}

After decades of theoretical and experimental efforts, quarkonium production remains a mystery. The lowest three $\UpsNS$ states in pp collisions provide an interesting probe of Quantum Chromodynamics (QCD). There are several quarkonium production models which predict different differential cross section shapes at high transverse momentum ($\pT$) in pp collisions. The non-relativistic QCD model(NRQCD) has adjustable parameters that are fit to data, which also impact the polarization predictions at all $\pT$~\cite{Cho1,Cho2}. In addition, both color single models (CSM) with additional higher order $\pT$ dependent corrections~\cite{Artoisenet} and the $k_{t}$-factorization model are consistent with data from the Large Hadron Collider (LHC)~\cite{Baranov}.

Different theoretical models also predict different polarizations of quarkonium. In the context of NRQCD, quarkonium are predicted to be transversely polarized, which disagree with the negligible polarization measured for the $\Jpsi$ by CDF~\cite{CDF_Jpsi}. The $\Upsilon$ satisfies the nonrelativistic approximation much better than the $\Jpsi$, making the $\Upsilon$ polarization a more decisive test of NRQCD, especially at large $\pT$. The measurement of the $\Jpsi$ polarization includes both directly produced $\Jpsi$ and those from feed-down decays of heavier charmonia. Given the absence of feed-down component, the measurement of the $\PsiTwo$ polarization should be particularly informative.

The $\UpsNS$ differential cross section times branching ratio to $\mu^{+}\mu^{-}$ integrated over $|y| < $ 0.6 in a given $\pT$ bin of width $\Delta\pT$,
\begin{equation}
\frac{d\sigma(pp \rightarrow \UpsNS)}{d\pT}|_{|y|<0.6} \times \mathcal{B} (\UpsNS \rightarrow \mu^{+}\mu^{-}) =
  \frac{N^{fit}_{\UpsNS}(\pT)}{L_{int} \cdot \Delta\pT \cdot \varepsilon(\pT) \cdot \mathcal{A}(\pT) } \,,
\end{equation}

Where $N^{fit}_{\UpsNS}$ is the number of $\UpsNS$ events in a $\pT$ bin of width $\Delta\pT$, $\varepsilon(\pT)$ is the dimuon efficiency, $L_{int}$ is the integrated luminosity, $\mathcal{A}(\pT)$ is the polarization-corrected acceptance.

The polarization of the $J^{PC} = 1^{--}$ quarkonium states can be measured through the study of the angular distribution of the leptons produced in their $\mu^{+}\mu^{-}$ decay~\cite{EPJC},
\begin{equation}
\label{eq:angular_distribution}
W(\costh,\varphi|\vec{\lambda}) \, \propto \,
\frac{1}{(3 + \lambda_{\vartheta})}(1  +  \lambda_{\vartheta} \cos^2 \vartheta\,
+ \lambda_{\varphi} \sin^2 \vartheta \cos 2 \varphi
+ \lambda_{\vartheta \varphi} \sin 2 \vartheta \cos \varphi ) \,,
\end{equation}

Where $\vartheta$ and $\varphi$ are the polar and azimuthal angles, respectively, of the $\mu^{+}$ with respect to the z axis of the chosen polarization frame. Improved experimental measurements of quarkonium polarization require measuring all the angular distribution parameters, $\vec{\lambda} = (\lamth, \lamph, \lamthph)$, in different polarization frames, as well as a frame-invariant polarization parameter, $\tilde{\lambda} = (\lambda_\vartheta + 3 \, \lambda_\varphi) / (1-\lambda_\varphi)$. Three popular frames are used: the centre-of-mass helicity (HX) frame, where the z axis coincides with the direction of the quarkonium momentum in the laboratory; the Collins-Soper (CS) frame, whose z axis is the bisector of the two beam directions in the quarkonium rest frame; and the perpendicular helicity (PX) frame, with the z axis orthogonal to that in the CS frame.

\section{CMS $\UpsNS$ differential cross section analysis}
In this analysis, a dimuon data sample with an integrated luminosity of 4.9 $fb^{-1}$ was used. The data was collected by the CMS experiment at the LHC in 2011 in proton-proton collisions at $\sqrt{s}$ = 7 TeV.

\begin{figure}[htb]
\centering
\includegraphics[height=2.5in]{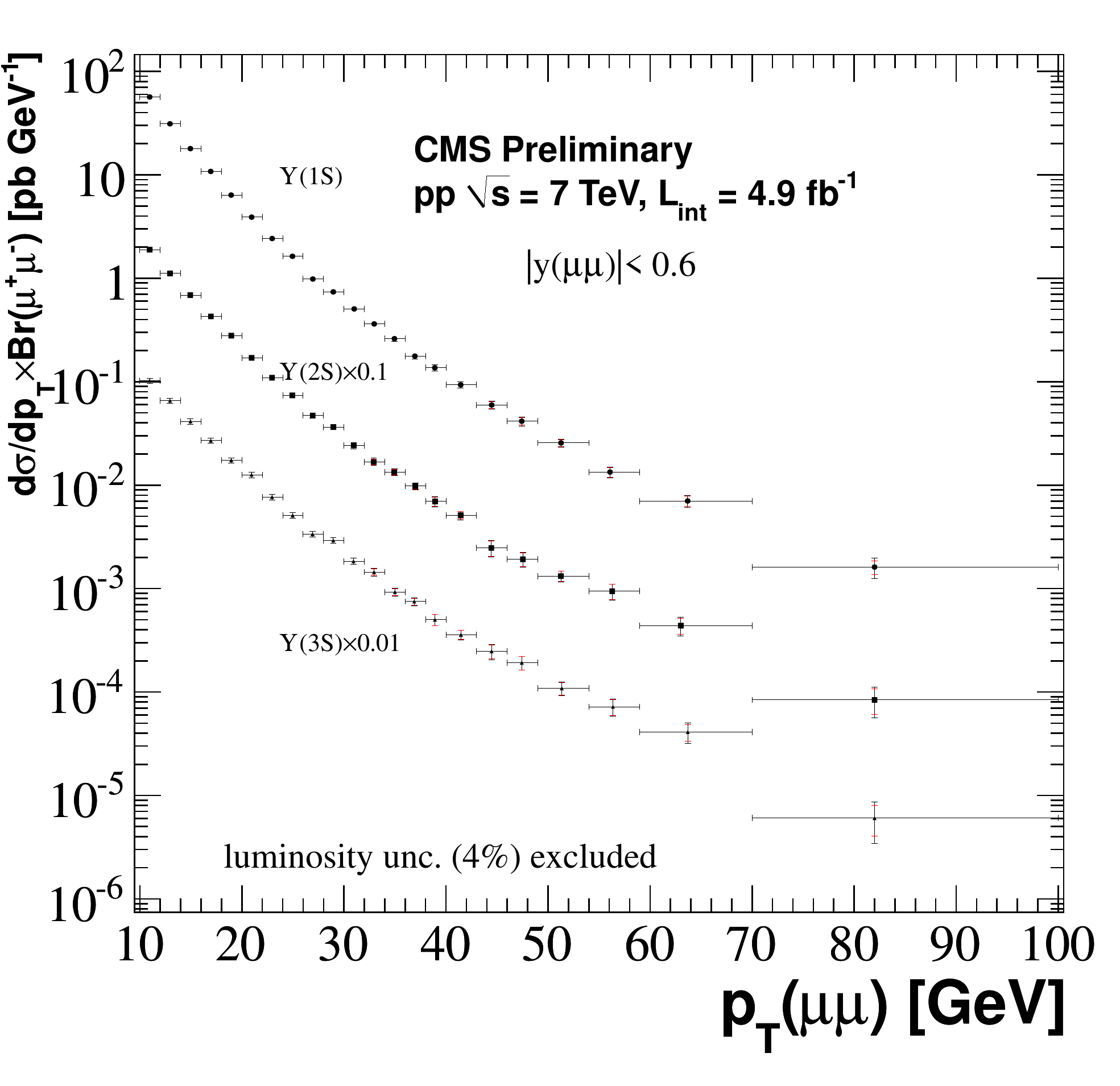}
\includegraphics[height=2.5in]{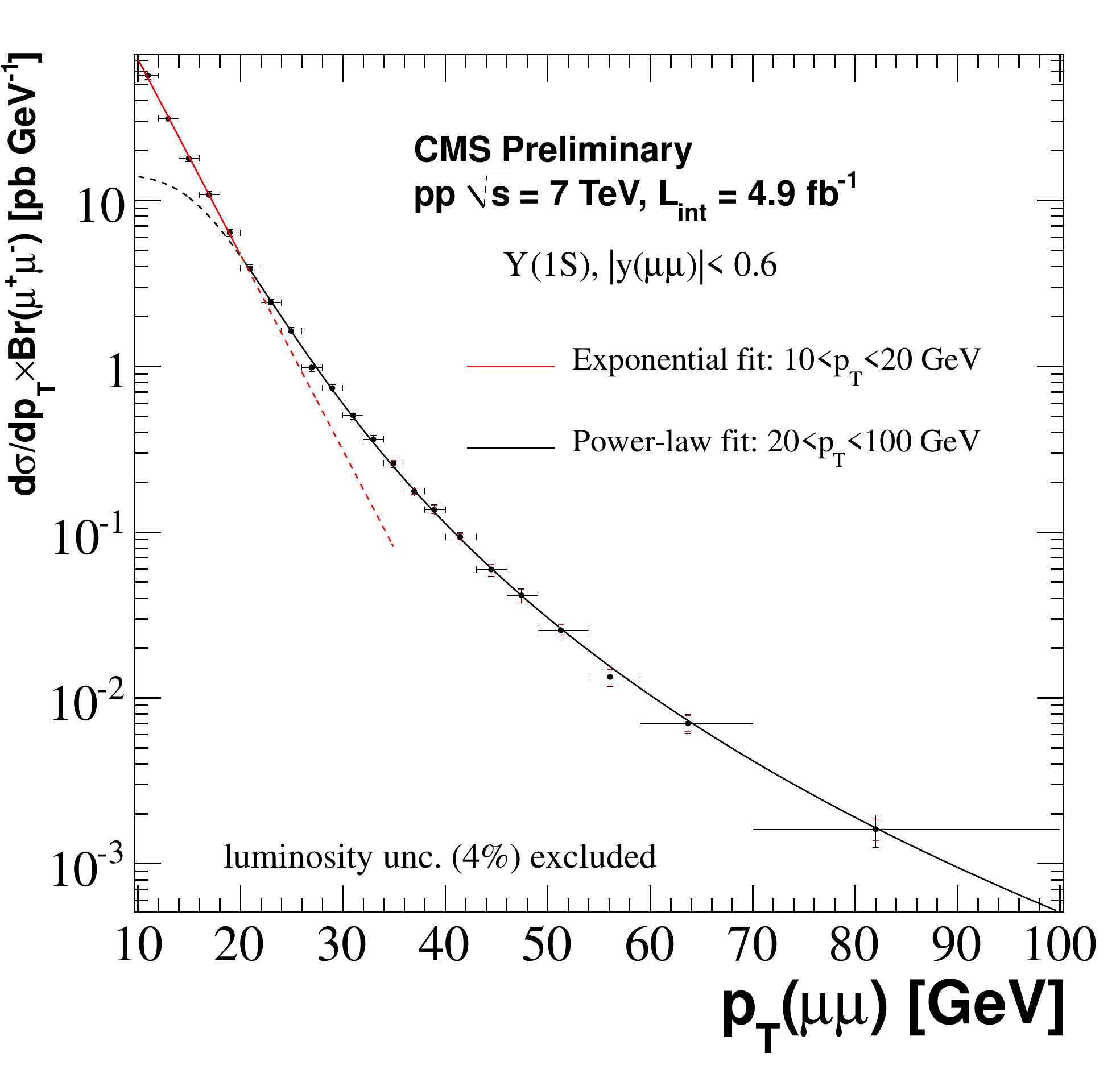}
\caption{Left: differential cross sections for each $\UpsNS$. The $\UpsTwo$ and $\UpsThree$ states are scaled by 0.1 and 0.01 respectively for display purposes. The total uncertainty is shown by the error bars including the uncertainty of the luminosity, and the smaller (red) error bars show the statistical uncertainty. Right: illustration of power-law behavior of the high-$\pT$ production. The red line shows an exponential fit to the data with 10 GeV $< \pT <$ 20 GeV.}
\label{fig:CrossSection}
\end{figure}

The $\UpsNS$ line-shape is determined directly from data, using the measured track momenta and their uncertainties, along with the exact mass distribution including final-state radiation (FSR) effects. Each dimuon event has a mass uncertainty computed from the track error matrices which is used to determine the dimuon invariant mass uncertainty distribution in each $\pT$ bin.

The differential cross sections are shown in Figure~\ref{fig:CrossSection}. The cross section is described by a simple exponential for 10 GeV $< \pT < 20$ GeV. For $\pT >$ 20 GeV the data lie above the exponential shape and the slope changes.

These new measurements of $\UpsNS$ production at $\sqrt{s}$ = 7 TeV provide precision differential cross sections for $\pT$ in the 10-100 GeV range. There is a transition from nearly exponential cross section decrease with $\pT$ to power-law behavior for all three $\UpsNS$ states. This suggestion of a change in the nature of the production process is mirrored by the $\pT$ dependence of the $\UpsTwo$ /$\UpsOne$  and $\UpsThree$ /$\UpsOne$ production ratios. The need to develop an understanding of $\UpsNS$ production mechanisms that would explain the
power-law behavior at high $\pT$ presents a challenge to theoretical models.

\section{CMS $\UpsNS$ and prompt $\Jpsi$ and $\PsiTwo$ polarization analysis}

CMS measured the polarizations of $\UpsNS$ and prompt $\Jpsi$ and $\PsiTwo$, using a dimuon data sample with an integrated luminosity of 4.9 $fb^{-1}$, collected by the CMS experiment at the LHC in 2011 in proton-proton collisions at $\sqrt{s}$ = 7 TeV. There are 222 k $\UpsOne$, 82 k $\UpsTwo$ and 51 k $\UpsThree$ with $\pT >$ 10 GeV and rapidity $|y| < $ 1.2 in this sample. For $\UpsNS$, the analysis is performed in 5 $\pT$ bins, with boundaries 10, 12, 16, 20, 30, and 50 GeV, and 2 $|y|$ ranges, $|y| < $ 0.6 and 0.6 $< |y| <$ 1.2. The total numbers of prompt plus nonprompt $\Jpsi$ events are 2.3 M and 2.4 M in the rapidity bins $|y| < $ 0.6 and 0.6 $< |y| <$ 1.2, respectively. The corresponding $\PsiTwo$ yields are 126 k, 136 k, and 55 k for $|y| < $ 0.6, 0.6 $< |y| <$ 1.2, and 1.2 $< |y| <$ 1.5, respectively. For $\PsiNS$, in each of these $|y|$ ranges, the analysis is performed in several $\pT$ bins, with boundaries at 14, 16, 18, 20, 22, 25, 30, 35, 40, 50, 98 and 70 GeV for the $\Jpsi$, and 14, 18, 22, 30, and 50 GeV for the $\PsiTwo$.

\begin{figure}[htb]
\centering
\includegraphics[height=1.7in]{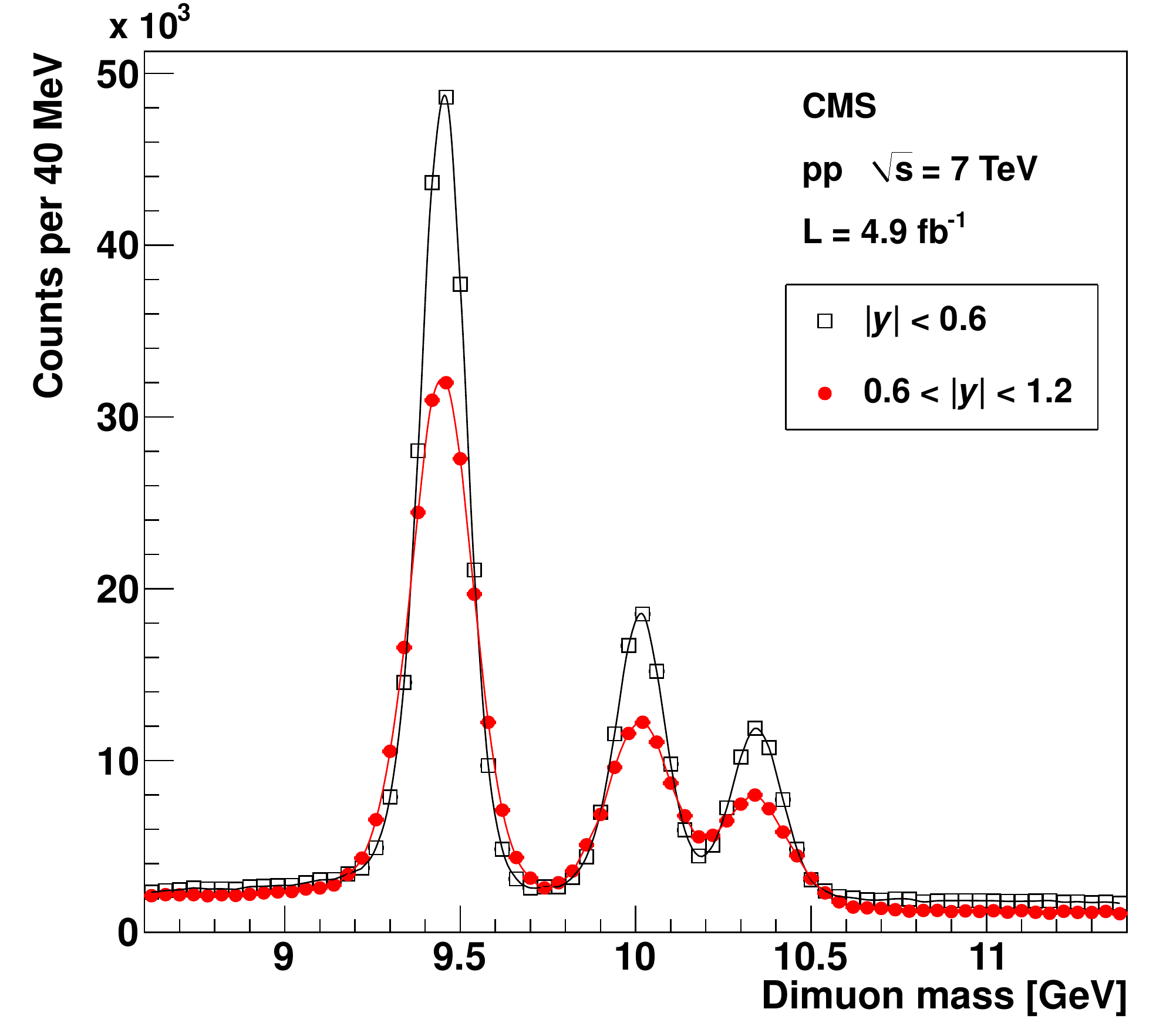}
\includegraphics[height=1.7in]{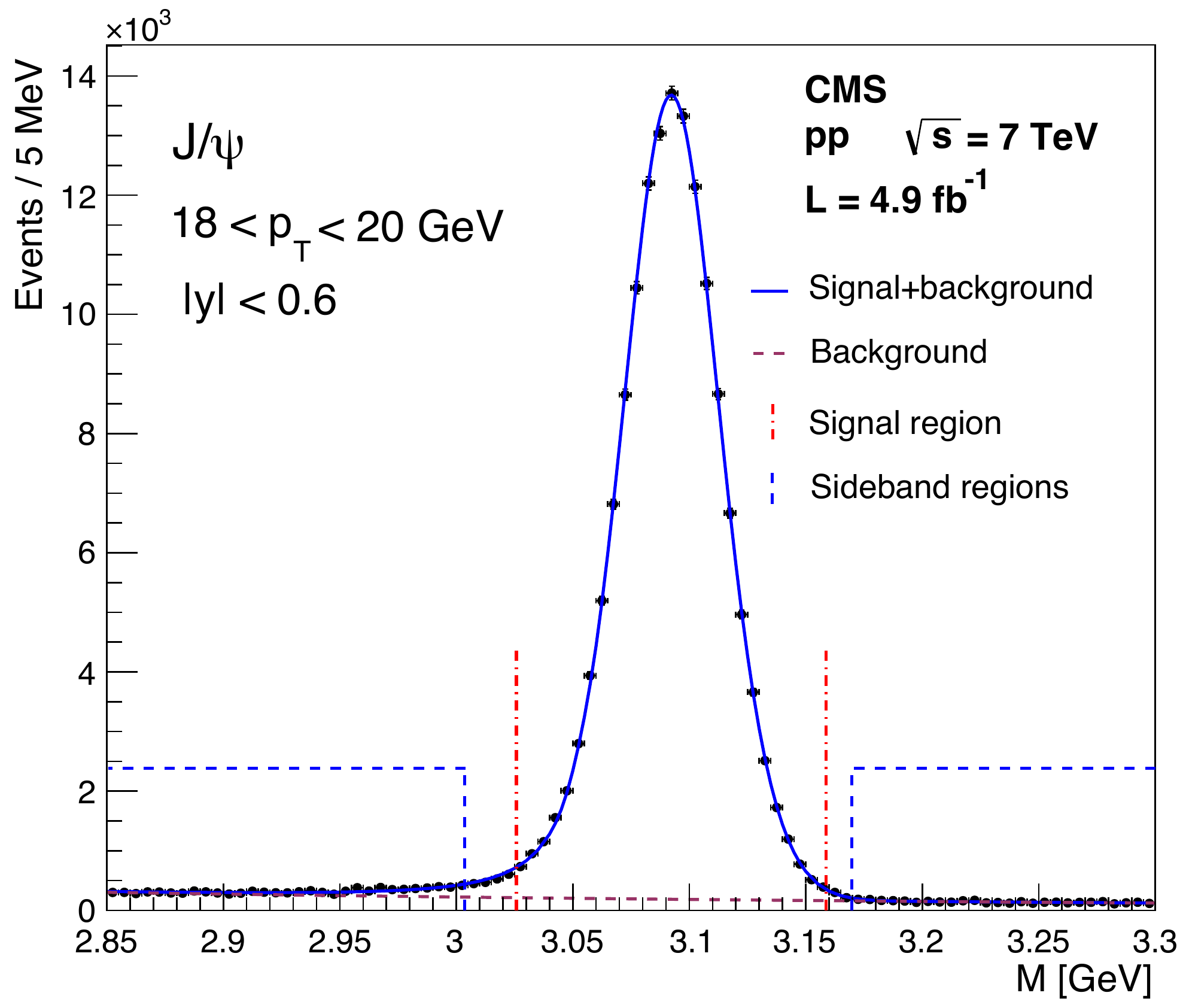}
\includegraphics[height=1.7in]{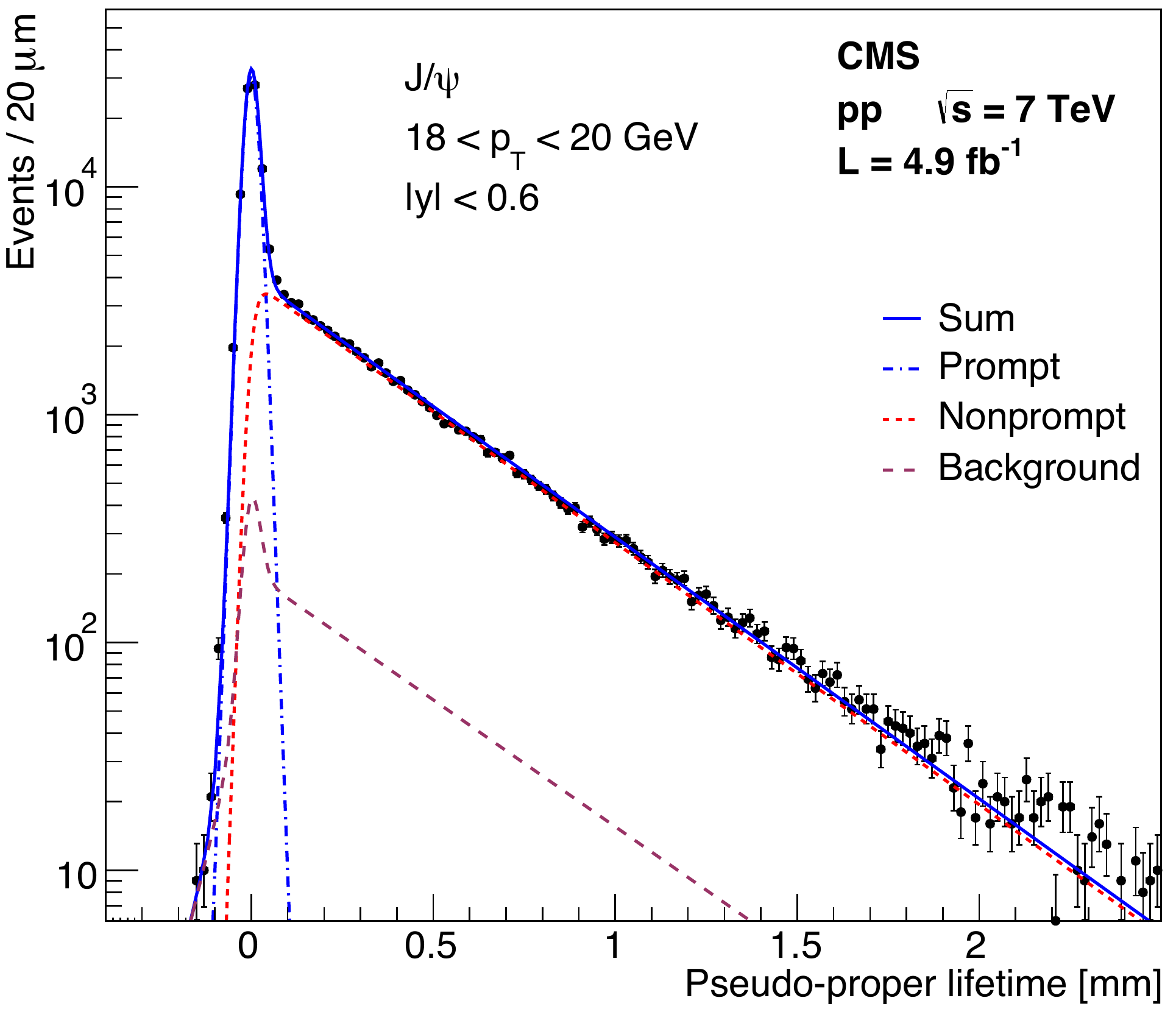}
\caption{ Left: dimuon mass distributions in the $\Upsilon$ region for $|y| < $ 0.6 (open squares) and 0.6 $< |y| <$ 1.2 (closed circles). Middle: dimuon mass distribution in the $\Jpsi$ region for an intermediate $\pT$ bin and $|y| < $ 0.6. Right: pseudo-proper lifetime distribution in the $\Jpsi$ region for an intermediate $\pT$ bin and $|y| < $ 0.6. }
\label{fig:masslifetime}
\end{figure}

As shown Figure~\ref{fig:masslifetime}, the dimuon mass distribution of $\UpsNS$ is well described by three Crystal-Ball functions representing the $\Upsilon$ peaks, and a second-degree polynomial function determined from the low- and high-mass sidebands. The dimuon mass distribution of $\Jpsi$ and $\PsiTwo$ is described by two Crystal-Ball functions representing each peak, and an exponential function representing the underlying continuum background.

To minimize the fraction of charmonia from B decays in the sample used for the polarization measurement, a prompt-signal region is defined using the dimuon pseudo-proper lifetime~\cite{CMS_Jpsi},  $\ell = L_{xy} \cdot m_{\psi(nS)} / \pT$, where $L_{xy}$ is the transverse decay length in the laboratory frame.
The resolution of pseudo-proper lifetime is modelled by the per-event uncertainty. The right figure of Figure~\ref{fig:masslifetime} shows pseudo-proper lifetime distribution for dimuons in the $\Jpsi$ signal regions, with the unbinned maximum-likehood fit result.

To extract the polarization parameters, the posterior probability distribution (PPD) is directly calculated. Firstly, events distributed according to the background model are removed from the data sample, based on a likelihood ratio criterion. The PPD for the average values of the polarization parameters ($\vec{\lambda}$) inside a particular kinematic cell is then defined as a product over the remaining events (i),

\begin{equation}
\label{eq:total_likelihood}
  \mathcal{P}(\vec{\lambda})= \prod_{i} \mathcal{E}(\vec{p}_1^{\,(i)},\vec{p}_2^{\,(i)})\,,
\end{equation}

where $\mathcal{E}$ is the probability distribution as a function of the two muon momenta $\vec{p}_{1,2}$ in event $i$. The numerical results and graphical representations are obtained from the 1D and 2D projections of the PPD.

The systematic uncertainties were studied on data and with pseudo-experiments based on simulated events. The final PPD of the polarization parameters is the average of the PPDs corresponding to all hypotheses considered in the determination of the systematic uncertainties. The total uncertainties are dominated by systematics at low $\pT$ and statistics at high $\pT$. The frame-invariant parameter $\lamtilde$ results obtained in the three frames (CS, HX, and PX) are in good agreement, as required in the absence of unaddressed systematic effects.

\begin{figure}[htb]
\centering
\includegraphics[height=3.in]{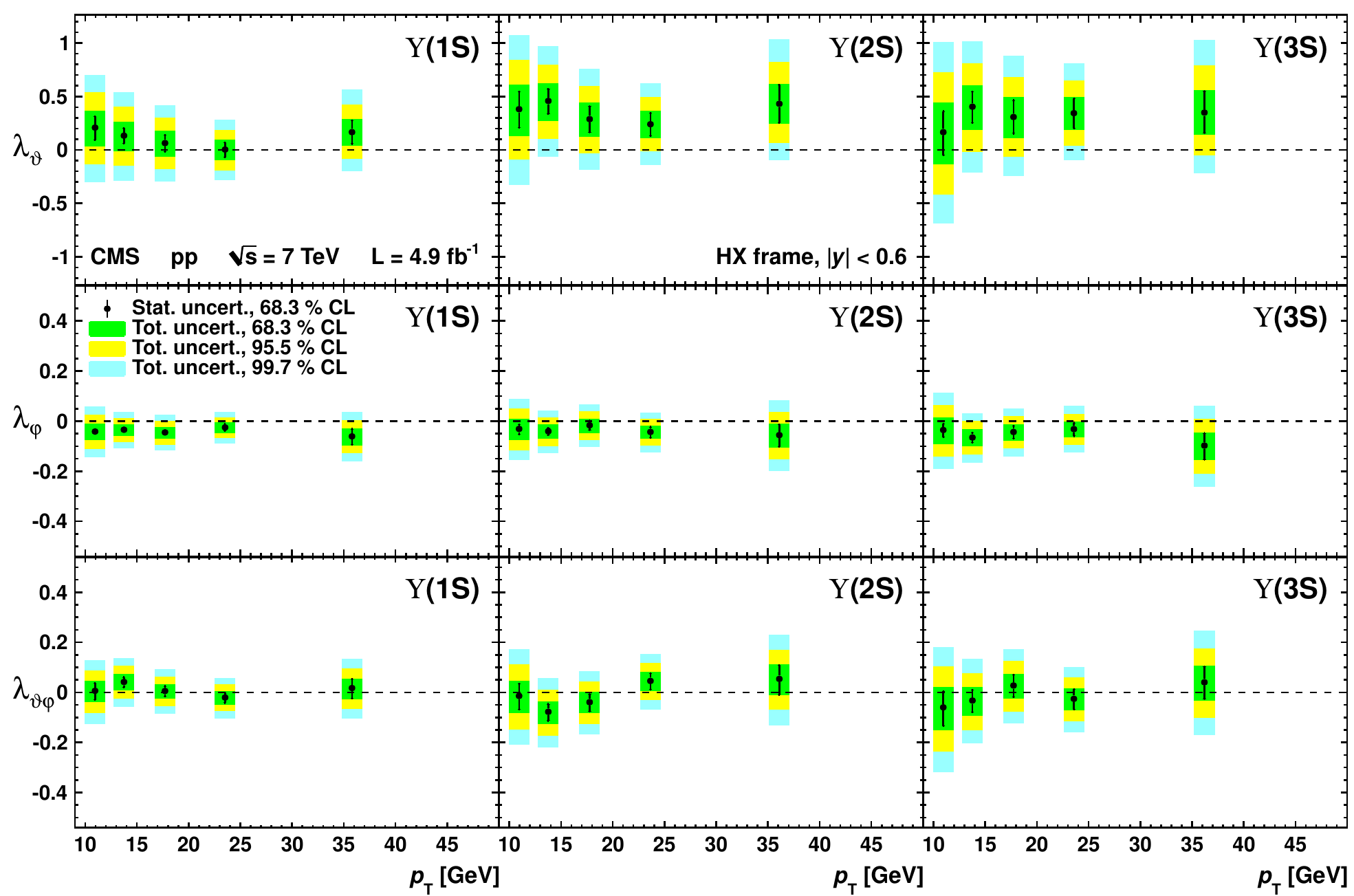}
\includegraphics[height=3.in]{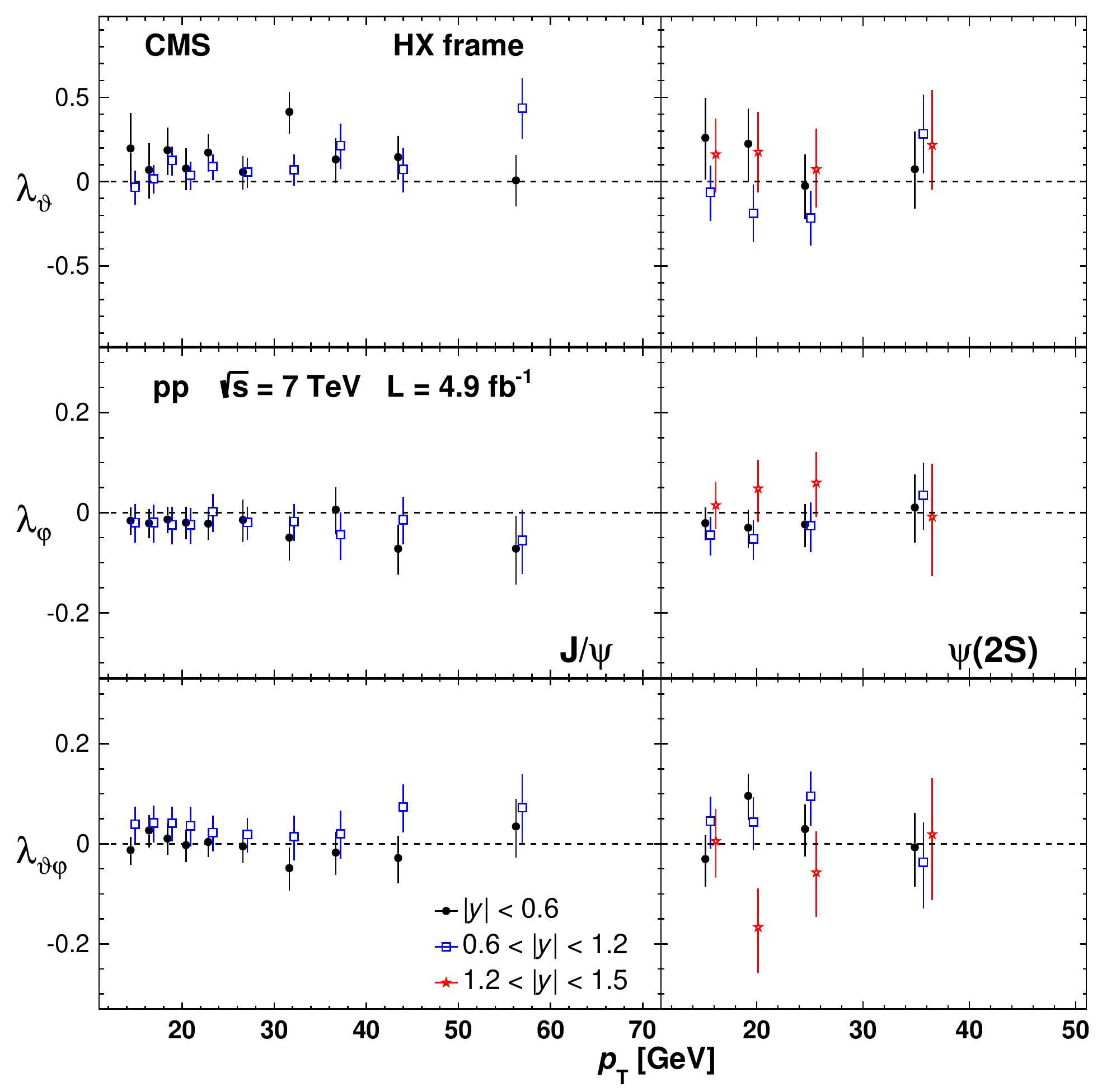}
\caption{Top: values of the $\lamth$ (top), $\lamph$ (middle), and $\lamthph$ (bottom) parameters for the $\UpsOne$ (left), $\UpsTwo$ (middle), and $\UpsThree$ (right), in the HX frame, as a function of the $\pT$ for $|y| <$ 0.6. The error bars indicate the 68.3\% CL interval when neglecting the systematic uncertainties. The three bands represent the 68.3\%, 95.5\%, and 99.7\% CL intervals of the total uncertainties.
Bottom: polarization parameters $\lamth$, $\lamph$, and $\lamthph$ measured in the HX frame for prompt $\Jpsi$ (left) and $\PsiTwo$ (right), as a function of $\pT$ and for several $|y|$ bins. The error bars represent total uncertainties at 68.3\% CL.}
\label{fig:lambda}
\end{figure}

The frame-dependent $\lambda$ parameters measured in the HX frame are presented, for $\UpsNS$ and prompt $\Jpsi$ and $\PsiTwo$, in Figure~\ref{fig:lambda}, as a function of $\pT$ and $|y|$. The results in CS and PX frames can be seen in~\cite{UpsPol,PsiPol}. None of the three polarization frames shows large polarizations, excluding the possibility that a significant polarization could remain undetected because of smearing effects induced by in appropriate frame choices.

\section{Conclusion}
The differential cross sections of $\UpsOne$, $\UpsTwo$, and $\UpsThree$ for 10 $< \pT <$ 100 GeV were measured by CMS using data collected in 2011, corresponding to an integrated luminosity of 4.9 $fb^{-1}$. A transition from exponential (low $\pT$) to power-law (high $\pT$) in the differential cross sections was observed. Three anisotropy parameters $\lamth$, $\lamph$, $\lamthph$ and the frame invariant parameter $\lamtilde$ were measured in three polarization frames (HX, CS and PX) for the $\UpsOne$, $\UpsTwo$, $\UpsThree$ and prompt $\Jpsi$ and $\PsiTwo$. All the measured $\lambda$ parameters are close to zero, excluding large polarizations in the explored kinematic regions. These results are in clear disagreement with existing next-to-leading order (NLO) NRQCD calculations and provide a good basis for significant improvements in the understanding of quarkonium production in high-energy hadron collisions.


\end{document}